\documentclass[twocolumn,showpacs,preprintnumbers,amsmath,amssymb,aps,superscriptaddress]{revtex4}

\usepackage{graphicx}
\usepackage{dcolumn}
\usepackage{bm}
\usepackage{latexsym}
\usepackage{amsmath}
\usepackage{dcolumn}
\usepackage{amsfonts}
\usepackage{subfigure}       \usepackage{indentfirst}
\begin{document}

\title{Dynamical scaling of imbibition in columnar geometries}

\author{M. Pradas}
\email{pradas@ecm.ub.es}

\affiliation{Departament d'Estructura i Constituents de la Mat\`eria\\
Universitat de Barcelona, Av. Diagonal, 647, E-08028 Barcelona,
Spain}
\author{A. Hern\'{a}ndez-Machado}

\affiliation{Departament d'Estructura i Constituents de la Mat\`eria\\
Universitat de Barcelona, Av. Diagonal, 647, E-08028 Barcelona,
Spain}

\author{M. A. Rodr\'{\i}guez}

\affiliation{Instituto de Fisica de Cantabria (CSIC-UC), Avda. los
Castros, E-39005 Santander, Spain}

\date{\today}

\begin{abstract}

Recent experiments of  imbibition in columnar geometries show
interfacial fluctuations whose dynamic scaling  is not compatible
with the usual non local model governed by surface tension that results from a macroscopic description. To
explore this discrepancy, we exhaustively  analyze numerical integrations of a
phase-field model with dichotomic columnar disorder. We find that two
 distinct behaviors are possible depending on the capillary
contrast between both values of disorder. In a high contrast case, where interface
evolution is mainly dominated by the disorder, an inherent anomalous scaling is always observed. Moreover, in agreement with experimental work, the interface motion has to be described through a local model.
On the other hand, in a lower contrast case, interface is dominated by interfacial tension and can be well modeled by a non local model. We have studied both spontaneous and forced-flow imbibition situations, giving a complete set of scaling exponent in each case, as well as, a comparison to the experimental results.

\end{abstract}


\maketitle

\section{Introduction}\label{Sec:Introduction}
Fluid--fluid displacements in porous media is a subject of much
interest in industrial processes and material characterization so as in
environmental problems ranging from petroleum recovery and
irrigation to retention of waste waters \cite{Sahimi-95,AL04}. We
restrict our analysis to the case in which the invading  fluid,
that wets preferentially the medium,  is more viscous than the
displaced resident fluid giving rise to compact rough interfaces.
This process can take place either spontaneously at constant
pressure, {\it spontaneous imbibition}, or by the application of an
external pressure at constant
injection rate, {\it forced-flow imbibition} \cite{AL04}.  The surface can be represented
by a single--valued function of position $\vec{x}$ and time $t$,
$h=h(\vec{x},t)$ \cite{Barabasi-Stanley}. In many cases, when
fluctuations have a thermal origin or can be reduced to that, the
interfacial fluctuations are self-affine and follow the dynamic
scaling of Family--Vicsek \cite{Family-Vicsek-1985}. Then, a complete description of the dynamical growth is possible with only two independent exponents. However, when disorder is relevant enough to interfere with the  geometry of the interface, a more generic scaling can apply,  and
one more independent exponent is necessary to reach the dynamical
description. It is the so-called anomalous scaling \cite{RA00}, and it has been observed in many different experimental and numerical situations during the last decade \cite{LE93,LO98,MA97,SO02b,SO03,SO05}.

Several experiments of imbibition in distinct geometries have been
proposed in the last years. There are experiments that use paper
as a disordered medium, \cite{Buldyrev-92-I,Horvath-95,Kwon-96,Zik-97,Balankin-2000} and
others performed in Hele--Shaw cells (two parallel glass
plates separated by a narrow distance) with a random
distribution of glass beads as a disordered medium
\cite{Rubio-89,Horvath-91,He-92}.  Using the same geometry, other
methods of generating a disordered medium have been explored,
including random variations in gap spacing produced by a
predesigned surface relief of the bottom plate
\cite{SO02b,SO03,SO05,SO02} or by
roughened plates \cite{Geromichalos-2002}. Focusing our attention
on columnar geometries we have presented in previous works \cite{SO02b,SO03} experimental studies of {\it forced-flow imbibition} in a Hele--Shaw cell with a columnar quenched disorder, produced by dichotomic variations in the thickness of the bottom plate. We found that the interfacial dynamics followed an intrinsic anomalous scaling with varying exponents that were incompatible with the expected results from the usual macroscopic model, that gives characteristic exponents of a non local  model. Instead, we obtained good agreement with an heuristic model of diffusively coupled columns presenting local interactions. The essential change of behavior, which pass to be  dominated by non local to local interactions, is obviously due to the persistence of the columnar disorder. However the detailed physical mechanism is not clear.

In the present paper we use numerical integrations of a phase-field model with columnar disorder to explore this behavior. By imposing the columnar geometry of the interface motion at the macroscopic model,  we analytically derive the heuristic local equation presented in Ref.\ \cite{SO02b}. In addition, we show how both local and non local behaviors are observed in the numerical model through the variation of a
parameter modeling the contrast between the different capillary values present in the system. The outline of the paper is as follows. In Sec.\ \ref{Sec:PFM} we
introduce the phase-field model and the interfacial equations
obtained in the sharp interface limit. In Sec.\ \ref{Sec:Scaling} we show the
notions of generic scaling used to characterize the interfacial
dynamics. Section \ref{Sec: LowC} is devoted to analyze the numerical results in the
low capillary contrast case for both spontaneous and forced-flow
imbibition situations. Section \ref{Sec:HighC} deals with the case of high capillary
contrast and its connection with local growth models. Finally, in
Sec.\  \ref{Sec:Conclusions} we discuss the physical contents of phase-field and
macroscopic models as well as their relevance to explain the experimental
results.

\section{The phase-field and macroscopic model}\label{Sec:PFM}
Our numerical results shall be performed using the so-called phase-field model \cite{DU99,EL01,Aurora-EPL-01}. Such a model is based on the introduction of an order parameter $\phi$ which can take two limit values, $\phi=\pm\phi_{e}$ representing the two phases liquid/air of the system. The phase-field dynamics is controlled by a conserved equation based on a Ginzburg-Landau formulation, $\partial\phi/\partial t=\nabla M\nabla\mu$, where $\mu=\delta\mathcal{F}/\delta\phi$ is the chemical potential and the free energy is given by the functional $\mathcal{F}[\phi ]=\int dr [V(\phi )+\frac{1}{2}(\epsilon
\nabla \phi )^2]$. The phase-field equation then reads as 
\begin{equation} \label{phasefield}
\frac{\partial\phi}{\partial t}=\bm{\nabla}
M\bm{\nabla}\Big[V'(\phi)-\epsilon^{2}\nabla^{2}\phi \Big],
\end{equation}
where $M$ is a parameter that is taken constant in the liquid
phase and zero in the air phase. $V(\phi)$ is a potential
taken as $V(\phi)=-\frac{1}{2} \phi ^{2} + \frac {1}{4} \phi
^{4}-\eta(r) \phi$ and defines two stable phases through the
double well potential; the destabilizing linear term 
accounts the effect of a capillary force that makes the interface
to advance. The effect of an inhomogeneous capillarity is added by using a
dichotomic capillary noise with the values:
\begin{equation} \label{eq:CapNoise}
\eta = \left\{\begin{array}{ll}
     \eta_{0} & \\
      & \\
     \frac{\eta_{0}}{1-\eta_{A}} &
                \end{array} \right. {}\\
\end{equation}
We consider a columnar disorder $\eta(x)$ defined by single tracks of lateral size $L_{d}$ distributed along the $x$-direction in such a way that tracks with the high disorder value $\eta=\eta_{0}/(1-\eta_{A})$ occupy 35\% of the system length $L$. This is the same kind of disorder reported in the experimental work of Ref.\ \cite{SO02b}. Note that, for a given disorder realization, tracks wider than $L_{d}$ are obtained when two or more unit tracks are placed adjacently. 

\subsection{The macroscopic description of imbibition}

The use of phase-field models to reproduce imbibition experiments is based on the ability to get the same results than those obtained from a macroscopic model. Indeed, in the so-called sharp interface limit $\epsilon\to 0$, a matched asymptotic expansion of the field $\phi$ around a kink solution of Eq.\ (\ref{phasefield}), $\phi_{0}=-\phi_{eq} tanh(w/\sqrt(2))$, can be performed, recovering then the basic macroscopic equations for the usual pressure $p$, velocity $v$,
interfacial curvature $\kappa$ and the columnar capillary disorder which we called $\eta(x)$:
\begin{eqnarray}
&\bm{v}=-K \bm{\nabla} p,{} \label{darcy}\\
&\nabla ^{2}p=0,{}\label{laplace}\\
&\Delta p_{int} =\sigma \kappa +\eta(x).{}\label{gibbs-thomson}
\end{eqnarray}
The macroscopic variables and parameters are defined from the phase-field formulation as \cite{HM03}:
$$ p=\phi_{eq} \mu_{1},\quad K=\frac{M}{2\phi_{eq}^2},\quad \sigma=\frac
{1}{2} \int\!\!\mathrm{d}w\Big(\frac{\partial \phi _{0}}{\partial w}\Big)^{2},$$
$\mu_{1}$ being the first order term of the expansion on
$\epsilon$ of the chemical potential and $\sigma$ playing the role of a interfacial tension.
The three equations of the macroscopic model are well known from
phenomenological arguments involving conservation laws. Darcy's law, Eq.\ (\ref{darcy}), arises from an averaging procedure of Navier-Stokes equations at low Reynolds number, when the geometry of the Hele-shaw cell is imposed. The Laplace equation, Eq.\ (\ref{laplace}), comes from imposing incompressibility of the liquid, and the Gibbs-Thomson relation, Eq.\ (\ref{gibbs-thomson}), comes from a principle of minimum interfacial energy. The capillary pressure at the interface can be expressed as $\eta(x)\sim 2\sigma\cos\theta/b$, where $\theta$ corresponds to the contact angle, and $b$ is the distance between the plates of Hele-Shaw cell. \\
In the experimental work reported in Ref.\ \cite{SO02b}, the random gap distribution is constructed by using  a fiberglass substrate with a pattern of copper tracks attached to the bottom plate of a Hele-Shaw cell. Tracks have a thickness $d$ with a lateral size of $L_{d}$  and are distributed along the lateral direction $x$ without overlap. Therefore, the gap of the Hele-Shaw cell has a dichotomic variation with two possible values $b$ and $b-d$.  Although in that case the capillarity is a 3-D effect of the cell, we can relate our numerical parameters of Eq.\ (\ref{eq:CapNoise}) as $\eta_{0}=1/b$ and $\eta_{A}=d/b$. Note that the parameter $\eta_{A}$ is related to the capillary contrast between both values of disorder. \\

\subsection {Spontaneous and forced-flow imbibition}

In our study, we will consider both situations of spontaneous and forced-flow imbibition by choosing conveniently the boundary conditions into the phase-field model \cite{LA05}. For spontaneous imbibition an applied constant pressure is imposed at the origin of the cell $\mu(x,y=0)=\mu_{a}$. In contrast, a pressure gradient has to be imposed at the origin, $K\partial_{y}\mu\vert_{y=0}=-V_{m}$, to reproduce forced-flow imbibition. Therefore, the main difference between both cases is found on the mean velocity of the interface. While in the forced-flow case, the interface evolves with the imposed constant velocity $V_{m}$, in the spontaneous imbibition case, the averaged interfacial height $H(t)$ follows the so-called Washburn's law $H(t)\sim t^{1/2}$. An exact expression of such an evolution can be obtained solving the following equation:
\begin{equation} \label{washburn}
\frac{dH}{dt}=\frac{K(\mu_{a}+\langle\eta\rangle)}{H(t)}
\end{equation}
that comes from (\ref{darcy}), (\ref{laplace}), and (\ref{gibbs-thomson}) with
$\kappa=0$ and $\langle\eta\rangle=\langle\eta(x)\rangle_{x}$. The expression for $H$ reads
$H(t)=\sqrt{H(0)^{2}+2at}$ where $a=K(\langle\eta\rangle+\mu_{a})$. Therefore, spontaneous imbibition has a slowing-down dynamics with a mean velocity $V_{m}(t)\sim t^{-1/2}$.\\

\subsection{Equations for the interface}

From the macroscopic equations, Eqns.\ (\ref{darcy}), (\ref{laplace}) and (\ref{gibbs-thomson}), it is possible to obtain an equation for the moving interface by means of a Green analysis. The Green function $G(r,r')$ in our case follows a Poisson equation with a unit source at $r'$, $\nabla^{2}G(r/r')=\delta(r-r')$ evaluated at the plane 2-D. The expression for the interface is obtained using the Green identity \cite{BA89}:
\begin{equation}\label{green}
\begin{array}{ll}
\int_{\Omega_{L}} \mathrm{d}\bm{r}' [p(\bm{r}')\nabla'^{2}
G(\bm{r}/\bm{r}')-G(\bm{r}/\bm{r}')\nabla'^{2} p(\bm{r}')]=   \\
& \\
\int_{S_{L}} \mathrm{d}\bm{s}'\cdot p(s')\bm{\nabla}'G(s,s')-\int_{S_{L}} \mathrm{d}\bm{s}'\cdot
G(s,s')\bm{\nabla}' p(s'),
\end{array}
\end{equation}
integrated over the volume of the liquid $\Omega_{L}=\{x,0\le y
\le h(x,t) \}$, $h(x,t)$ being the interfacial position. Depending
on the intensity of the quenched noise two cases are relevant in
our study.

\subsubsection{Low capillary contrast. Linearized equations}

We assume that the quenched noise does not impose any special geometry and the interface can be linearized around its averaged value, $h(x,t)=H(t)+\delta h(x,t)$, with $H(t)=\langle h(x,t)\rangle_{x}$. Keeping the first order on $\delta h$ fluctuations in Eq.\ (\ref{green}), and imposing Darcy's law, Eq.\ (\ref{darcy}), at the interface boundary condition, one can obtain the linearized interface equation \cite{AL04,DU99,LA05}, which in Fourier space reads as
\begin{equation}\label{eq:Lineal}
\delta\dot{\tilde{h}}_{k}=-\sigma K\vert k\vert
k^{2}\delta\tilde{h}_{k}-V_{m}\vert k\vert\delta\tilde{h}_{k} +K \vert
k\vert\eta_{k},
\end{equation}
where we have supposed that correlations do not grow up faster in time than the mean height of the interface, $\vert k\vert H(t)\gg 1$. Under this limit, interface fluctuations follow the same  Eq.\ (\ref{eq:Lineal}) both for spontaneous and forced-flow imbibition \cite{LA05,Aurora-EPL-01,PA03}. However, it is worth to mention here that the presence of Washburn's law ($V_{m}(t)\sim t^{-1/2}$) in spontaneous imbibition give rise to dynamic crossover lengths and therefore, as it has been pointed out in Ref. \cite{Pradas07}, a rich variety of different scaling regimes can be observed. The crossover length scale can be explicitly seen as a balance between the surface tension term $\sigma K\vert k\vert^{3}$ and the drift term $\vert k\vert V_{m}$ \cite{DU99}
\begin{equation}\label{eq:dube}
\xi_{\times}=2\pi \Bigg(\frac{\sigma K
}{V_{m}}\Bigg)^{1/2}.
\end{equation}
In the forced-flow case, $V_{m}$ is constant and thus the crossover length is just a static length scale separating two different regimes. However, as it has been observed in several numerical results \cite{DU99,LA05,intrinsic-superrough}, this crossover length acts as a cutoff for the interface fluctuations growth due to the interface being asymptotically flat on length scales larger than $\xi_{\times}$.
\begin{figure}[h]
\centering
\includegraphics[width=0.3\textwidth,origin=c,angle=0]{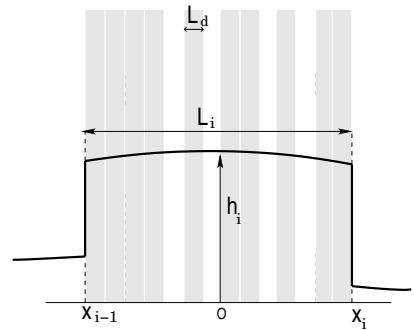}
\caption{Scheme of the interface $h_{i}(t)$ advancing through  an effective channel $L_{i}$ composed of several unit tracks $L_{d}$. The gray stripes correspond to  tracks with a high capillary noise,  $\eta=\eta_{0}/(1-\eta_{A})$, and the white ones to  tracks with a low capillary noise,  $\eta=\eta_{0}$.}\label{fig:Channel}
\end{figure}

\subsubsection{High capillary contrast. Coupled channel equations}
\begin{figure*}[!]
\centering
\subfigure[Forced-flow imbibition]{\includegraphics[width=0.4\textwidth,origin=c,angle=0]{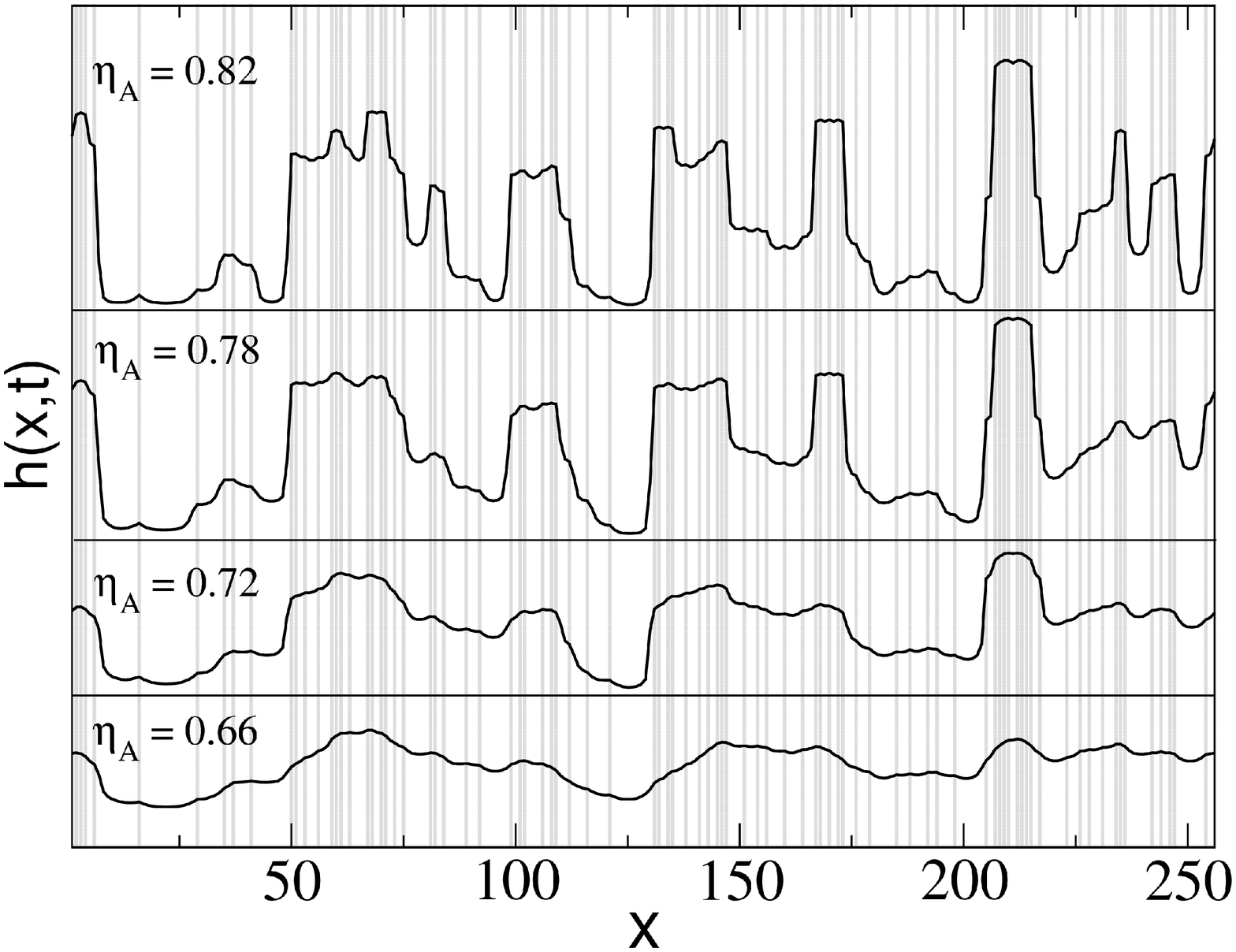}}\label{fig:intAV}
\hspace{0.5cm}
\subfigure[Spontaneous imbibition]{\includegraphics[width=0.4\textwidth,origin=c,angle=0]{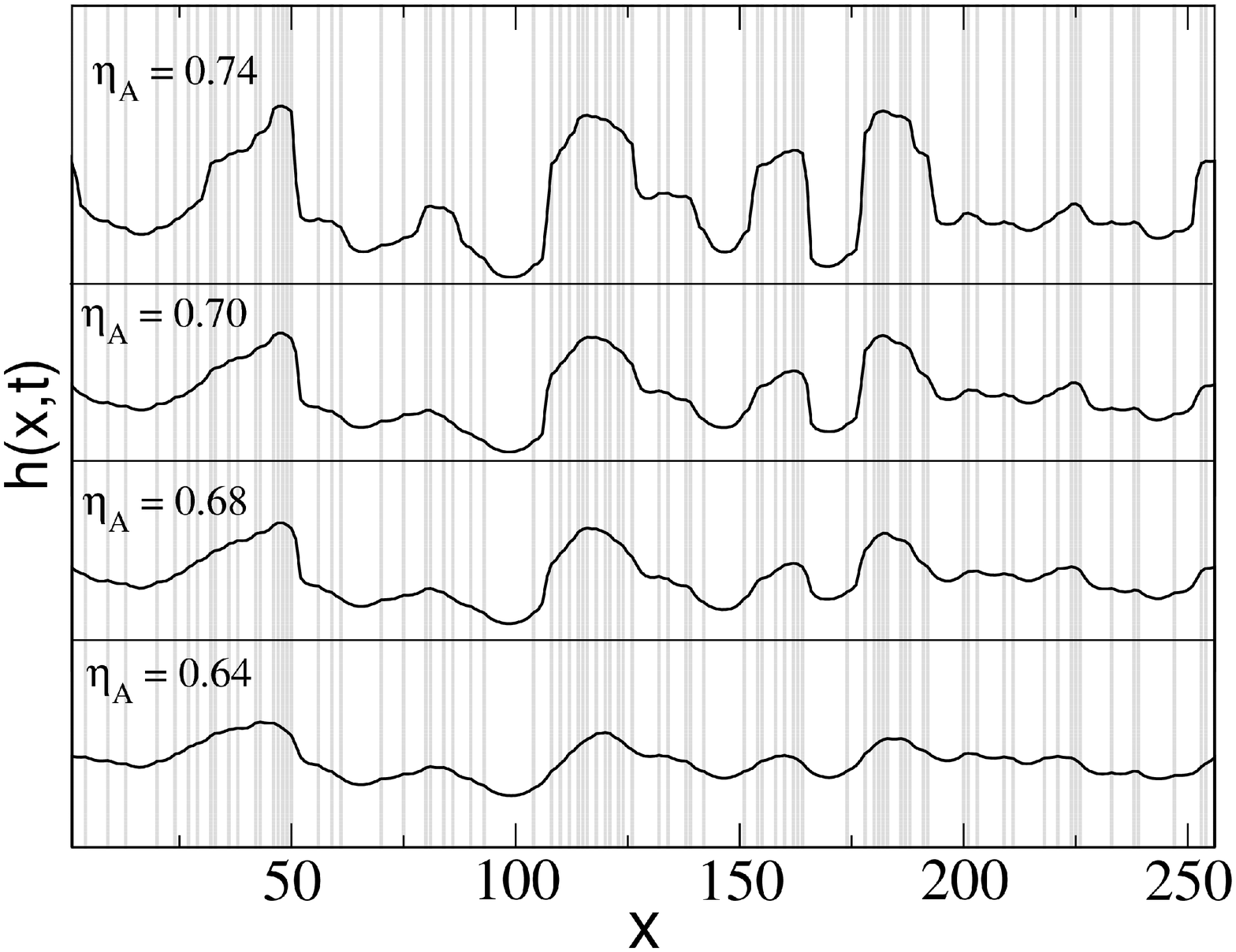}}\label{fig:intAW}
\caption{Interface profiles at equal times but different values of capillary contrast $\eta_{A}$.  The columnar disorder is also plotted: gray tracks are points where the dichotomic disorder takes its highest value of Eq.\ (\ref{eq:CapNoise}).}\label{fig:IntA}
\end{figure*}

The situation becomes quite different when the capillary contrast is increased. In the experimental work of Ref.\ \cite{SO02b}, where forced-flow imbibition is studied for high capillary contrast, it is observed that the interface motion can be modeled through a phenomenological local equation. Our purpose now is to derive such an equation directly from the macroscopic model. To do this, we  shall assume that the noise is so strong that the interface adopts the columnar geometry of the disorder. The procedure consists then on integrating Eq.\ (\ref{green}) in a closed surface  along the profile of the \emph{i-th} effective channel. Here, an effective channel, or simply channel $i$, is defined as a group of single adjacent tracks of which the majority has the same disorder value, in such a way that the interface advances as a compact surface through each channel. An example is depicted in Fig.\ \ref{fig:Channel}, where the channel is composed of several tracks of which the majority  has a high disorder value (gray tracks) and therefore, the averaged disorder of the channel $\eta_{i}$ is larger than the mean disorder of the whole system $\langle\eta\rangle$. Note that the surrounding channels $i-1$ and $i+1$ must have a mean disorder value $\eta_{i}<\langle\eta\rangle$. The width of the channel is defined as $L_{i}$. A numerical example of these channels can be seen in Fig.\ 2. 

We are considering the forced-flow imbibition case with the following boundary conditions at the top and the bottom of the channel:
\begin{eqnarray}
&\partial_{y}p(x,y)\vert_{y=h_{i}}=-\frac{1}{K}\dot{h}_{i}(t){}\\
&\partial_{y}p(x,y)\vert_{y=0}=-\frac{V_{m}}{K},{}\label{BCgradp}\\
& p(x,0)\sim \frac{V_{m}H(t)}{K}-\langle\eta\rangle,{}
\end{eqnarray}
where the pressure at the origin has been estimated by using the relation $\frac{P_{int}-p(x,0)}{H(t)}=-\frac{V_m}{K}$, which comes essentially from Eq.\ (\ref{BCgradp}), taking that $P_{int}=-\langle\eta\rangle$ is the mean capillary pressure of the whole system. It means that the pressure at the origin is changing on time in order to get a mean constant velocity for the whole interface. Using these boundary conditions, and taking the general expression for the two-dimensional Green function
$$G(x-x',y-y')=-\frac{1}{(2\pi)^2}\iint\mathrm{d}\bm{k}\frac{e^{i(x-x')k_{x}}\cdot e^{i(y-y')k_{y}}}{k_{x}^{2}+k_{y}^{2}}, $$
we can evaluate the different boundary integrals of the top and bottom segments of Eq.\ (\ref{green}) as
\begin{align*}
\int_{x_{i-1}}^{x_{i}}\!\!\!\mathrm{d}x\ \bm{n}\cdot \big[ p(x,y)\bm{\nabla}\hat{G}(x,y)\big]_{y=h_{i}}&=0 {} \\
\int_{x_{i-1}}^{x_{i}}\!\!\!\mathrm{d}x\ \bm{n}\cdot\big[\hat{G}(x,y)\bm{\nabla}p(x,y)\big]_{y=h_{i}}&=\frac{L_{i}C}{\pi K}\dot{h}_{i}(t), {} \\
\int_{x_{i}}^{x_{i-1}}\!\!\!\mathrm{d}x\ \bm{n}\cdot \big[p(x,y)\bm{\nabla}\hat{G}(x,y)\big]_{y=0}&=\frac{V_{m}H(t)}{2K}-\frac{\langle\eta\rangle}{2}, {} \\
\int_{x_{i}}^{x_{i-1}}\!\!\!\mathrm{d}x\ \bm{n}\cdot\big[\hat{G}(x,y)\bm{\nabla}p(x,y)\big]_{y=0}&=\frac{V_{m}h_{i}(t)}{2K}-\frac{L_{i}C}{\pi K}V_{m}, {}
\end{align*}
where $C=\int_{\lambda}^{\infty}\!\mathrm{d}u\frac{sin(u/2)}{u^{2}}$, with $\lambda=L_{i}/L$ being a \emph{cut off} due to  the finite size $L$ of the system. We have also supposed enough wide channels to ensure that $L_{i}>h_{i}(t)$, which means that we are taking the initial times, before interface gets saturated. 
The nomenclature used for the Green function means that it is evaluated at the interface, $\hat{G}(x,y)\equiv G(0,h_{i}/x,y)$. Therefore, Eq.\ (\ref{green}) can be written as
\begin{align}\label{channel}
\frac12 p(0,h_{i})&=\int_{h_{i}}^{0}\!\!\!\mathrm{d}y'\Big[p(x,y')\partial_{x}\hat{G}(x,y')-\hat{G}(x,y')\partial_{x} p(x,y)\Big]_{x_{i}}\nonumber \\ -\int_{0}^{h_{i}}&\!\!\!\mathrm{d}y'\Big[p(x,y')\partial_{x}\hat{G}(x,y')-\hat{G}(x,y')\partial_{x} p(x,y)\Big]_{x_{i-1}}\nonumber\\ 
-a_{i}&\dot{h}_{i}(t)+a_{i}V_{m}-\frac{\langle\eta\rangle}{2}.
\end{align}
The two first terms of the right-hand side are due to the flow between neighbor channels. We have defined the parameter $a_{i}\equiv\frac{L_{i}C}{\pi K}$. In order to get an equation for the time evolution of the interface $h_{i}(t)$, we define the following coupling coefficients between channels as the ratio between channel flow and height differences 
\begin{align*}
D_{i}\equiv&\frac{\int_{h_{i}}^{0}\!\!\!\mathrm{d}y'\Big[p(x,y')\partial_{x}\hat{G}(x,y')-\hat{G}(x,y')\partial_{x} p(x,y')\Big]_{x_{i}}}{a_{i}(h_{i+1}-h_{i})}, \\%
D_{i-1}\equiv&\frac{\int_{0}^{h_{i}}\!\!\!\mathrm{d}y'\Big[p(x,y')\partial_{x}\hat{G}(x,y')-\hat{G}(x,y')\partial_{x} p(x,y')\Big]_{x_{i-1}}}{a_{i}(h_{i}-h_{i-1})}.%
\end{align*}
The coupling variable $D_i$ has to be understood as a diffusion coefficient which depends on each channel $i$. As a general case, it may be taken as a random variable. Moreover, we shall assume that  $D_i$ does not vary in time  during the initial times, before interface gets saturated.\\
Then, we can write Eq.\ (\ref{channel}) as an inhomogeneous diffusion equation between channels:
\begin{equation}\label{difchanel1}
\dot{h_{i}}(t)=\bm{\nabla} D_{i}\bm{\nabla} h_{i}+V_{m}+ \frac{1}{2a_{i}}\Big(-p_{i}(t)
-\langle\eta\rangle\Big).
\end{equation}
where $p_{i}(t)\equiv p(0,h_{i})$ corresponds to the pressure at the interface. We are assuming that the pressure at the interface is time-dependent, which is based on the experimental results reported in Ref. \cite{SO02b}. In such experimental work, the local velocity of the interface at each channel follows an expression similar to Washburn's law  until it reaches the saturation value $V_{m}$. In order to take into account such behavior into the equation, we consider that the pressure at the interface can be expressed as $p_{i}(t)\sim -\Delta p_{int}-cu_{i}(t)$, where the pressure difference $\Delta p_{int}$  is given by the usual Gibbs-Thomson relation, Eq.\ (5), taking a negligible atmospheric pressure. The term $cu_{i}(t)$ is an effective kinetic term due to the local capillary forces at each channel, being $c$ an arbritrary constant, and can be explained in terms of the mass conservation. When the interface goes through a channel $i$ of a high capillary disorder ($\eta_{i}>\langle\eta\rangle$), its local velocity tends to initially increase up to a nominal value. In contrast, since we are imposing a constant velocity for the whole interface, the local velocity at the neighbor channel $i+1$ with a lower capillary disorder tends to decrease down to a nominal value. After reaching such nominal value in both cases, the local velocity decrease or increase asymptotically to the saturation value following the Washburn's behavior due to the capillary forces of each channel. Therefore, we are taking that $u_{i}(t)\sim \xi_{i}t^{-1/2}$ for  $t>0$, where $\xi_{i}$ is  a random variable defined as $\xi_{i}=(\eta_{i}-\langle\eta\rangle)/\vert \eta_{i}-\langle\eta\rangle\vert$, which takes the values $\xi_{i}=+1$ at the channel with the highest capillary value ($\eta_{i}>\langle\eta\rangle$), and $\xi_{i}=-1$ at the channel with the lowest capillary value ($\eta_{i}<\langle\eta\rangle$). In addition, we also suppose that the curvature of the interface can be approximated as a constant value $\kappa_{0}$, and only its sign depends on each channel as $\kappa_{i}\sim \xi_{i}\kappa_{0}$.\\ 
Then, rewriting the last term on the right-hand side of Eq.\ (\ref{difchanel1}), we get the final expression
\begin{equation}\label{difchanel}
\dot{h_{i}}(t)=\bm{\nabla} D_{i}\bm{\nabla} h_{i}+V_{m}+ \xi_{i}\Big(\bar{v}_{i}+\bar{a}_{i}t^{-1/2}\Big),
\end{equation}
where we have defined the new constants as $\bar{v}_{i}=(\vert \eta_{i}-\langle\eta\rangle\vert+\sigma\kappa_{0})/2a_{i}$, and $\bar{a}_{i}=\sigma c/2a_{i}$.\\
We thus conclude that  when the capillary disorder is large enough, the columnar geometry of the system leads to a local description for the interface motion.

\section{Generic scaling laws and the characterization of the interfacial dynamics} \label{Sec:Scaling}
Rough interfaces grow exhibiting power laws in both a horizontal correlation length $\ell_{c}\sim t^{1/z}$, that accounts the range of correlation, and a vertical growth length, like the interface width $W(t,L)\sim t^\beta$. $W$ is defined as the deviation of the height $h(x,t)$ as $W(L,t)=\langle (h(x,t)-\overline{h(x,t)})^2\rangle^{1/2}$ (where $\langle...\rangle$ and $\overline{...}$ mean sample and spatial average respectively), $z$ and $\beta$ are the so-called dynamic and growth exponents, which completely characterize the growth of self-affine processes. The saturation of the surface occurs in a saturation time $t_{s}$, when the correlation length reaches the system length $\ell_{c}(t_{s})=L$. Above this time, interfacial width scales as $W(t>t_{s},L) \sim L^{\alpha}$, $\alpha$ being the roughness exponent which it is related to the others exponents through the scaling relation $\beta =\alpha /z$. To study local growth, we shall define local widths averaged on windows of size $\ell<L$, $w(\ell,t)=\langle (\overline {
h(x,t)-\overline{h(x,t)}_{\ell})^2}_{\ell}\rangle ^{1/2}$. Local growth is also given as a power law $ w(\ell,t)\sim \ell_{c}(t)^{\alpha}$ for $\ell>\ell_c$ and $w(\ell,t)\sim \ell_{c}(t)^{\alpha-\alpha_{loc}}$ for $\ell<\ell_{c}$, where  $ \alpha_{loc}$ is the local roughness exponent. In the case of self-affine growth, global scaling coincides with local scaling, $\alpha_{loc}=\alpha$, and the fluctuating interface is well characterized by only two independent exponents. However, following the method shown in Ref.\ \cite{RA00}, we must use at least three independent exponents  in a more general case. In our study we use time variation of the local width for several window lengths: $ w(\ell,t)\sim t^{\beta} g(\ell t^{-1/z})$, with
\begin{equation} \label{scalingwl}
g(u) = \left\{\begin{array}{ll}
     u^{\alpha_{loc}}, u\ll 1 & \\
      & \\
     const, u\gg1  &
                \end{array} \right. {}\\
\end{equation}
obtaining direct measures of the exponents $\beta$, and $\beta^{*}=\beta-\frac{\alpha_{loc}}{z}$. Indirect measures of $z$ and $\alpha$ can be obtained through a collapse of the individual figures as $w(\ell,t)\ell^{-\alpha}\sim g'(\ell t^{-1/z})$, where now the scaling function goes as $u^{\alpha_{loc}-\alpha}$ for $u<<1$ and $u^{-\alpha}$ for $u>>1$. Finally, although the collapse gives an indirect measure of exponents in a very robust way it is worth to have a direct measure of at least three exponents. Thus we also use the evolution of the power spectral density $S(k,t)=\langle |h(k,t)|^{2} \rangle$ that scales as $S(k,t)=\frac{1}{k^{2\alpha+1}} s(kt^{1/z})$, with
\begin{equation} \label{scalingsp}
s(u) = \left\{\begin{array}{ll}
     u^{2\alpha +1}, u\ll 1 & \\
      & \\
     u^{2(\alpha-\alpha_{s})}, u\gg 1  &
                \end{array} \right. {}\\
\end{equation}
Depicting $S(k,t)$ at different times $t$, we have the direct measure of the spectral exponent $\alpha_{s}$ that coincides either with the global roughness exponent $\alpha$, when power spectrum do not shift in time (self-affine or superrough scalings) or with the local one $\alpha_{loc}$, when a temporal shift is observed (intrinsic anomalous scaling).

\section{Low capillary contrast} \label{Sec: LowC}
\begin{figure}[!]
\centering
\includegraphics[width=0.4\textwidth,origin=c,angle=0]{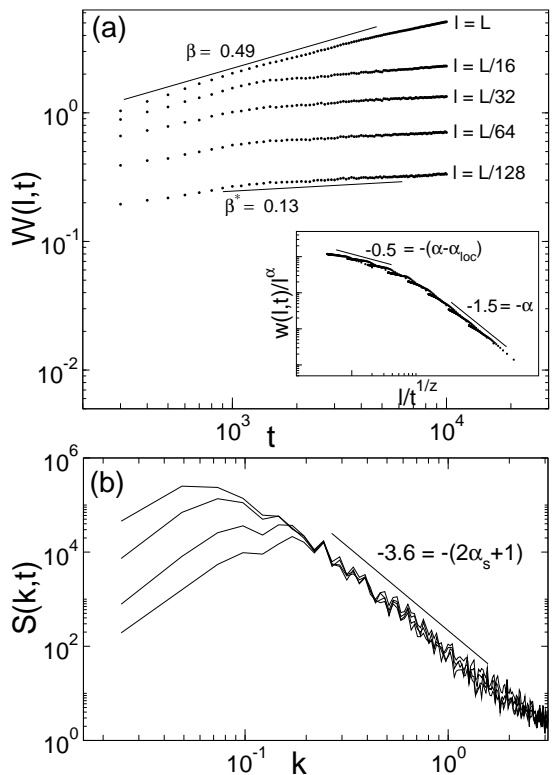}
\caption{Statistical analysis of the interface fluctuations in the forced-flow case with a low capillary contrast, $\eta_{A}=0.66$. (a) Local width $w(\ell,t)$ evaluated at different window sizes. The global and local growth exponents can be measured directly from the data. The inset shows the best data collapse for the scaling function using the values of $z=3.0$ and $\alpha=1.5$. It also suggests a local roughness exponent of $\alpha_{loc}=1.0$. (b) Interface power spectrum calculated at different times. It shows a roughness exponent of $\alpha_{s}=1.3\pm 0.2$.}\label{fig:StatZ3V}
\end{figure}
In this section we deal with the case of low capillary contrast in both forced-flow and spontaneous imbibition cases. It is worth to mention here that the numerical parameters used in all numerical results have been $\epsilon=1$, $M=1$ (dimensionless units), and $\eta_{0}=0.3$. Equation (\ref{phasefield}) has been integrated over a system of lateral size $L=256$ using a spatial grid of $\Delta x=1.0$ and a time step of $\Delta t=0.01$. The minimum length of the track disorder has been taken as $L_{d}=2$.

\subsection{Forced-flow imbibition}
We start to study a regime of low capillary pressures. The used value for the capillary contrast has been of $\eta_{A}=0.66$ and the mean velocity has been fixed to $V_{m}=0.0025$. The shape of the interface for a given realization is shown in Fig.\ \ref{fig:IntA}a, case $\eta_{A}=0.66$. We can see a smooth interface which is slightly correlated with the disorder. The results of the roughness analysis are shown in Fig.\ \ref{fig:StatZ3V}. From the local width $w(\ell,t)$ computed at different window sizes we can obtain the global and local growth exponents, $\beta=0.49\pm 0.05$ and $\beta^{*}=0.13\pm 0.05$. The best collapse of theses curves (shown at the inset of Fig.\ \ref{fig:StatZ3V}a is obtained tuning the values of $z=3.0$ and $\alpha=1.5$. The slopes of the scaling function agree with the previously calculated exponents suggesting also a value for the local roughness exponent of $\alpha_{loc}=1$ which is corroborated at the power spectrum calculated at different times. Since there is not any temporal shift between the lines of the power spectrum, we can assume that the interface fluctuations are described within the superrough anomalous scaling, and therefore $\alpha_{loc}=1$.  In addition, the spectral roughness exponent obtained from the power spectrum $\alpha_{s}=1.3\pm 0.2$ corresponds to the global roughness exponent, which is in agreement with the value obtained previously. These measured exponents are compatible with those obtained by the linear Eq.\ (\ref{eq:Lineal}) with a constant velocity $V_{m}$. Rescaling such a linear equation by the transformation $x\to bx, t\to b^{z}t, h\to b^{\alpha}h$, we have trivially that $z=3$ and $\alpha= 1.5$, and assuming superroughness ($\alpha_{loc}=1$), we get the remainder exponents as $\beta=\alpha/z=0.5$, $\beta^{*}=\beta-\frac{\alpha_{loc}}{z}=0.17$, in agreement with the measured exponents. Therefore, we can conclude that this regime is well modelled by the non local and linear equation (\ref{eq:Lineal}), taking into account only the surface tension regime.\\
If we increase the mean velocity of the interface, then the static crossover length, Eq.\ (9), decreases, obtaining that interface fluctuations saturate earlier, at the time when the correlation length $\ell_{c}$ reaches the crossover length, $t_{s}\sim \xi_{\times}^{z}$ \cite{AL04,LA05,intrinsic-superrough}.

\subsection{Spontaneous imbibition}
In spontaneous imbibition, the crossover length scale, Eq.\ (\ref{eq:dube}), becomes a dynamical scale and different regimes can be observed depending on the velocity of the interface \cite{intrinsic-superrough}. For low velocities, the initial correlation length $\ell_{c}\sim t^{1/z}$ is below the crossover length meaning that the relevant mechanism to damping the interface fluctuations is the surface tension with the characteristic dynamical exponent of $z=3$. On the other hand, for higher velocities, the crossover length acts as an effective correlation length of the interface fluctuations, giving rise to the genuine exponent of $z=4$. In order to study both regimes, we have controlled the initial velocity of the interface by choosing the initial height of the interface conveniently. We impose an initial height of $H(0)=199$ to study a low velocity regime, whereas a higher velocity regime will be achieved by simply putting $H(0)=1$.
\begin{figure}[!]
\centering
 \includegraphics[width=0.4\textwidth,origin=c,angle=0]{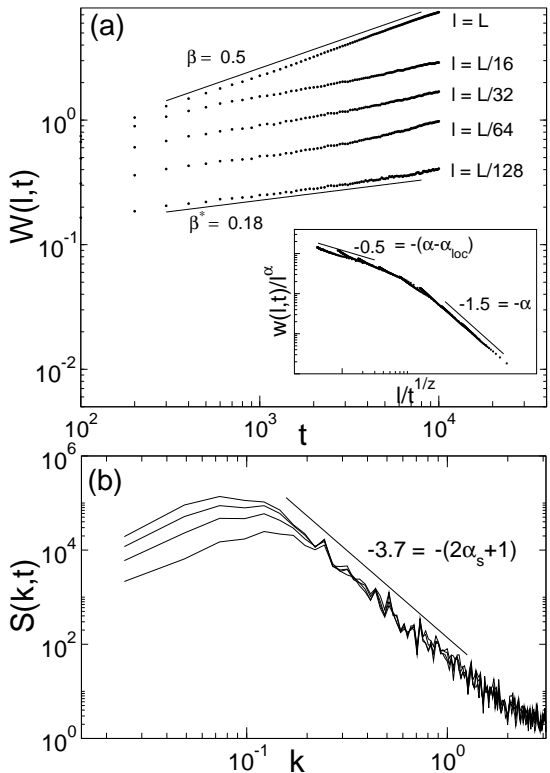}%
\caption{Statistical analysis of the interface fluctuations in the spontaneous case within a low velocity regime, and with a low capillary contrast, $\eta_{A}=0.64$. (a) Local width $w(\ell,t)$ evaluated at different window sizes. The global and local growth exponents can be measured directly from the data. The inset shows the best data collapse for the scaling function using the values of $z=3.0$ and $\alpha=1.5$. It also suggests a local roughness exponent of $\alpha_{loc}=1.0$. (b) Interface power spectrum calculated at different times. It shows a roughness exponent of $\alpha_{s}=1.35\pm 0.2$.}\label{fig:StatZ3W}
\end{figure}
\begin{figure}[!]
\centering
\includegraphics[width=0.4\textwidth,origin=c,angle=0]{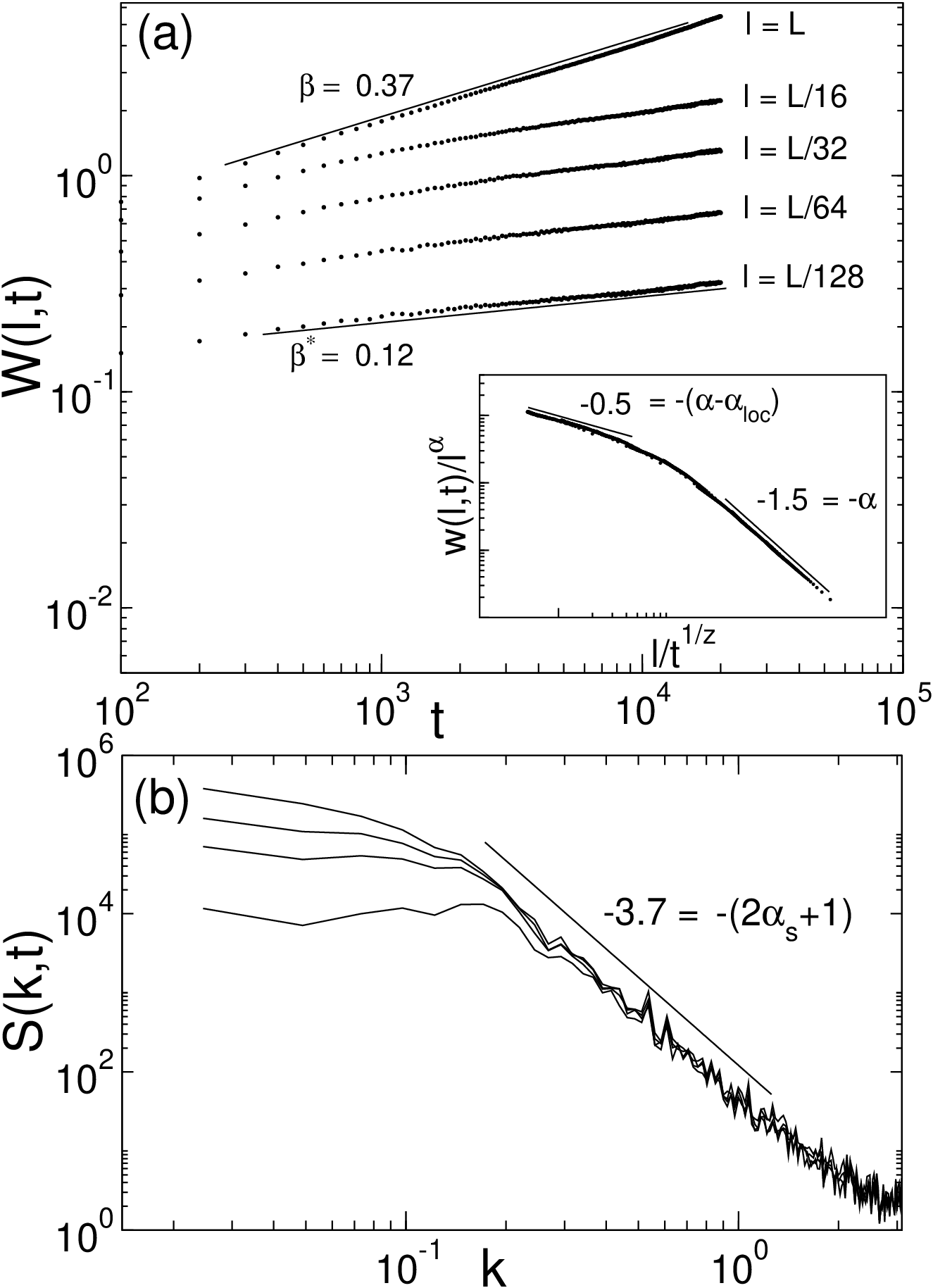}%
\caption{Statistical analysis of the interface fluctuations in the spontaneous case within a high velocity regime, and with a low capillary contrast, $\eta_{A}=0.64$. (a) Local width $w(\ell,t)$ evaluated at different window sizes. The global and local growth exponents can be measured directly from the data. The inset shows the best data collapse for the scaling function using the values of $z=4.0$ and $\alpha=1.5$. It also suggests a local roughness exponent of $\alpha_{loc}=1.0$. (b) Interface power spectrum calculated at different times. It shows a roughness exponent of $\alpha_{s}=1.35\pm 0.2$.\\}\label{fig:StatZ4W}
\end{figure}

\subsubsection{Low velocity regime}
Typical shapes of interfaces in spontaneous imbibition are depicted in Fig.\ \ref{fig:IntA}b. In the low capillary regime, case of $\eta_{A}=0.64$, the interface is weakly correlated with disorder. A roughness analysis shown in Fig.\ \ref{fig:StatZ3W} gives the exponents $\beta=0.5\pm 0.04$ and $\beta^{*}=0.18\pm 0.04 $ from a direct measure of the growing local width. The best data collapse on these figures provides the exponents $z=3.0$ and $\alpha=1.5$. Likewise, the slope of the scaling function gives $\alpha_{loc}=1$, which is corroborated by the power spectrum shown in Fig.\ \ref{fig:StatZ3W}b , where we get a spectral roughness exponent of $\alpha=1.35\pm 0.2$, without temporal shift between the curves, indicating we are dealing with a superrough scaling. Hence, the measured exponents are the same than those obtained in the forced-flow case. It was actually expected, since the relevant terms of Eq.\ (\ref{eq:Lineal}) at low velocities are the same in both cases. 

\subsubsection{High velocity regime}
When the initial interface velocity is increased, the velocity-dependent term of Eq.\ (\ref{eq:Lineal}) starts to be relevant and the new regime adopts the dynamics of the crossover length $\xi_{\times}\sim t^{1/4}$, getting then the expected dynamical exponent $z=4$ \cite{DU99}, keeping the spatial structure with the same roughness exponent as before, $\alpha =1.5$. The numerical results are presented in Fig.\ \ref{fig:StatZ4W}, obtaining $\beta=0.37\pm 0.03$, $\beta^{*}=0.12\pm 0.03$ and $\alpha_{s}=1.35\pm 0.2$ from direct measures, and $\alpha=1.5$, $z=4.0$, $\alpha_{loc}=1$ from the data collapse of the local widths, which are also in agreement with the linear description of Eq.\ (\ref{eq:Lineal}).

\section{High capillary contrast} \label{Sec:HighC}
\begin{figure}[!]
\centering
\includegraphics[width=0.4\textwidth,origin=c,angle=0]{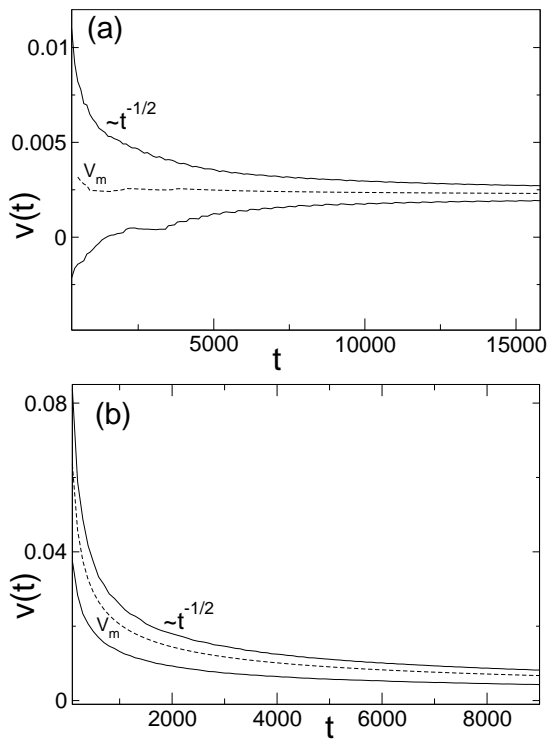}
\caption{Interface velocity $v(t)$ evaluated at two different points $x$ of the system (solid lines) and the mean interface velocity $V_{m}$ (dashed line) in both imbibition cases: (a) Forced-flow imbibition with $V_{m}=0.0025$, and (b) spontaneous imbibition within a high velocity regime. The curve above the mean velocity correspond to a channel with a high capillary disorder value (gray tracks in Fig.\ 2) and the curve below the mean velocity correspond to a channel with a low capillary disorder value (white tracks in Fig.\ 2). }\label{fig:Vmit}
\end{figure}
When the capillary contrast is increased the shapes of interfaces become sharper in both cases of imbibition, as  it can be seen in Fig.\ \ref{fig:IntA}. The effect of disorder is so strong that a kind of columnar geometry is also imposed on the interface. As we will see, two important points of the dynamics of interface fluctuations arise as effect of increasing the capillary contrast. First, interface motion seems to be described by local effects instead of the non local effects observed in the case of low capillary contrast. Second, scaling type changes from superrough to anomalous intrinsic. As in the case of low contrast, forced-flow and spontaneous imbibition cases present distinct patterns of fluctuations and they need to be studied separately.

\subsection{Forced-flow imbibition}
\begin{figure}[!]
\centering
\includegraphics[width=0.4\textwidth,origin=c,angle=0]{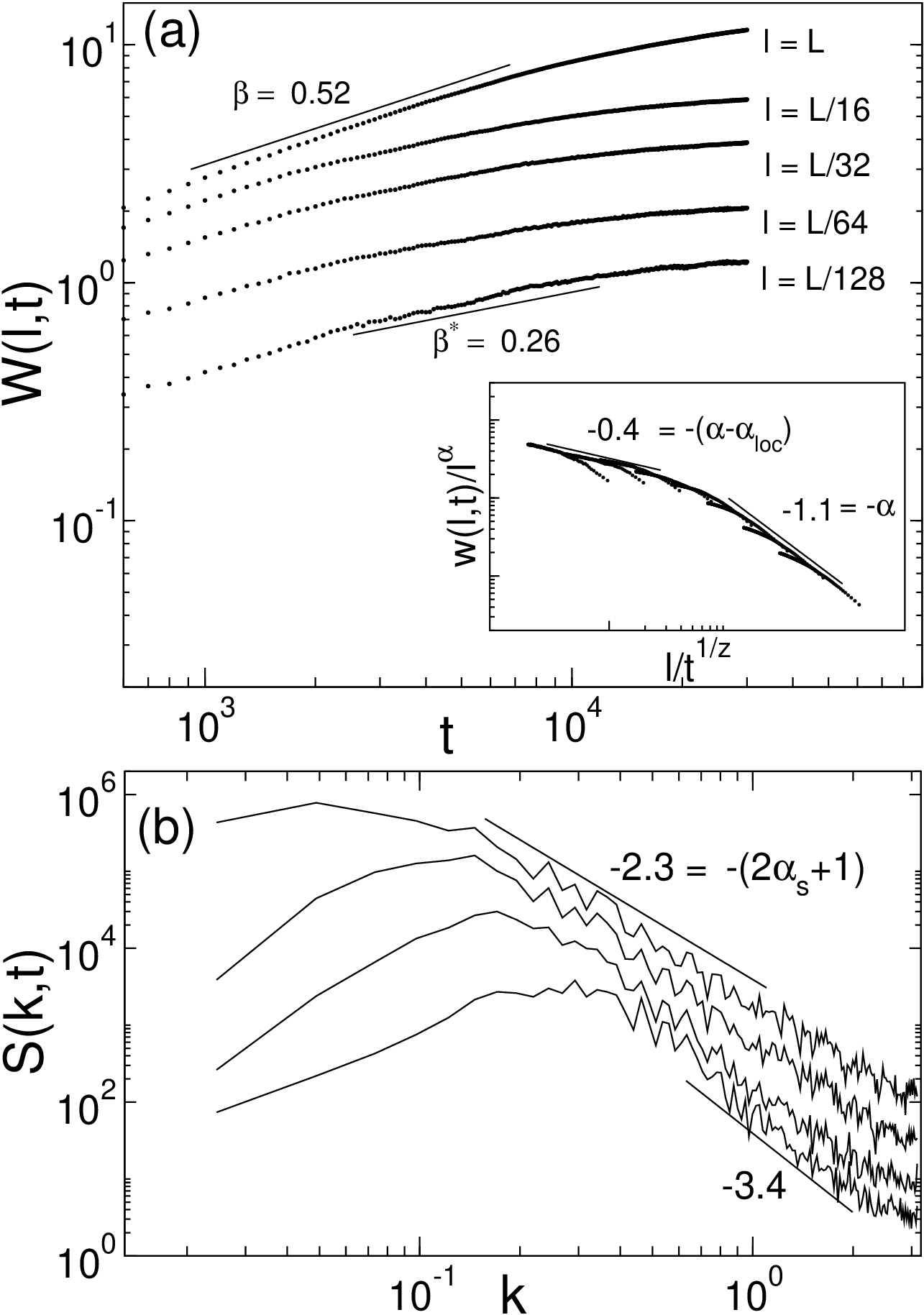}%
\caption{Statistical analysis of the interface fluctuations in the forced-flow case with a high capillary contrast, $\eta_{A}=0.72$. (a) Local width $w(\ell,t)$ evaluated at different window sizes. The global and local growth exponents can be measured directly from the data. The inset shows the best data collapse for the scaling function using the values of $z=2.2$ and $\alpha=1.1$. It also suggests a local roughness exponent of $\alpha_{loc}=0.7$. (b) Interface power spectrum calculated at different times. It shows a roughness exponent of $\alpha_{s}=0.65\pm 0.2$.\\}\label{fig:StatZ2V}
\end{figure}
We start to study the case of a capillary contrast $\eta_{A}=0.72$ (see Fig.\ \ref{fig:IntA}a). In Fig.\ \ref{fig:Vmit}a there are plotted the local interface velocity at two different points x of the system (solid lines) and the mean velocity $V_{m}$ (dashed line). The curve above the mean velocity corresponds to a channel $i$ with a high disorder value, that is $\eta_{i} > \langle\eta\rangle$, being $\eta_{i}$ the disorder of the channel $i$ and $\langle\eta\rangle$ the mean disorder of the whole system. The curve below the mean velocity corresponds to the next channel $i+1$ with a lower disorder value, that is $\eta_{i+1} < \langle\eta\rangle$. We can see that both profiles can be locally described by Eq.\ (\ref{difchanel}), $v_{\pm}\sim V_{m}\pm(\bar{v}+\bar{a}t^{-1/2})$, where $+$ and $-$ mean the channel with $\eta_{i}- \langle\eta\rangle>0$ and $\eta_{i}- \langle\eta\rangle<0$ respectively. Since the averaged velocity of the interface is low enough, the velocity $v_{-}$ can be initially negative \cite{SO02b}.\\
As in the previous section we are interested in calculating the different scaling exponents. The scaling analysis is shown in Fig.\ \ref{fig:StatZ2V}. From  direct measure of the interfacial local width slopes we get $\beta=0.52\pm 0.05$, and $\beta^{*}=0.26\pm 0.06$. The best data collapse is obtained using $\alpha=1.1$ and $z=2.2$. The power spectrum evolution changes with respect to the low contrast case since now there is a temporal shift between the curves, indicating the presence of inherent anomalous scaling. Hence, the spectral roughness exponent $\alpha_{s}=0.65\pm 0.2$ must be interpreted as the local roughness exponent, which is in agreement with the slopes of the collapsed scaling function. Note that these exponents are very close to the experimental exponents reported in Ref. \cite{SO02b}, $\beta=0.5\pm 0.04$, $\beta^{*}=0.25\pm 0.03$, $\alpha=1.0\pm 0.1$. As shown in this reference, the shape of the interfaces and the analysis of fluctuations are well reproduced by Eq.\ (\ref{difchanel}), which takes into account strong diffusive coupled channels. Therefore, we can conclude that simple numerical integrations of the phase-field model reproduce both the shape and the scaling analysis of interfaces.

\subsubsection{From low to high capillary contrast}
\begin{figure}[!]
\centering
\includegraphics[width=0.4\textwidth,origin=c,angle=0]{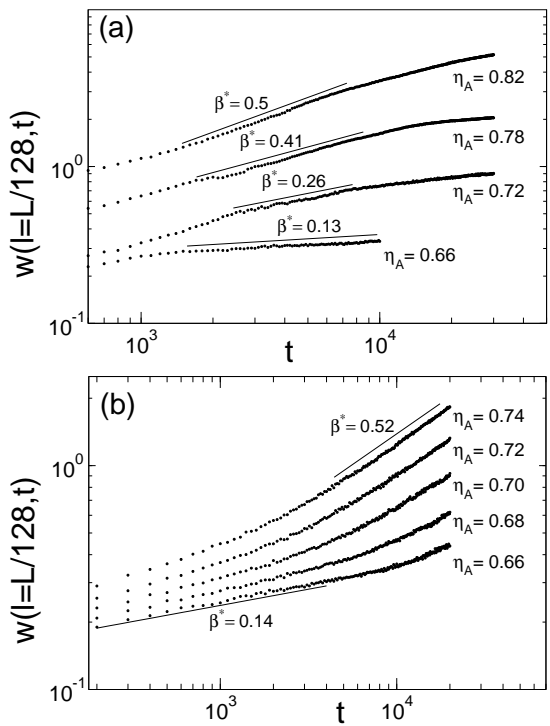}
\caption{Interface local width calculated in a small window of size $\ell=L/128$ for different capillary contrasts. The local growth exponent $\beta^{*}$ can be obtained directly from the data. (a) Forced-flow imbibition, $V_{m}=0.0025$. (b) Spontaneous imbibition within a high velocity regime.}\label{fig:wA}
\end{figure}
When we increase even more the values of the capillary contrast, interfaces become more correlated to the columnar disorder presenting quantitative changes on the scaling behavior. For instance, in Fig.\ \ref{fig:wA}a there is plotted the local width computed in a small window of length $\ell=L/128$. It allows us to calculate the local growth exponent $\beta^{*}$ for different capillary contrasts. The complete set of scaling exponents is presented in Table (\ref{tab}). We can see that for high capillary contrasts ($\eta_{A}>0.7$), interface fluctuations are always described by intrinsic anomalous scaling. In addition, the dynamics of the correlation is subdiffusive, ranging from $z=2$ to $z\!\to\!\infty$ in the highest contrast value. We must interprete this  extreme case as having fluctuations completely decoupled. It is characterized by the same local and global growth exponent $\beta=\beta^{*}=0.5$, which implies a dynamic exponent of $z=\infty$, and it can be understood as a regime where the correlation length $\ell_{c}\sim t^{1/z}$ does not grow in time anymore. Experimentally, it has also been observed  in the work carried out in Ref.\ \cite{SO03}. It is worth noting that there are two exponents $\beta=0.5$ and $\alpha_{loc}=0.5$ that remain constant.\\
Alternatively to the phase-field model results, this phenomenon can be reproduced by Eq.\ (\ref{difchanel})  taking a random diffusion coefficient $D(x)$ \cite{LO95}. By choosing a probability density $P(D)=N_{a}D^{-a}f_{c}(D/D_{max})$, $f_{c}$ being a \emph{cut off} function, Eq.\ (\ref{difchanel}) predicts a growth exponent $\beta=\alpha_{loc}=0.5$ independently of the $a$ value, a roughness exponent $1\le \alpha<\infty$, and a dynamic exponent $2\le z<\infty$, which is in accordance with the numerical values obtained using the phase-field model (see Table \ref{tab}). In this sense, we can say that the diffusion coefficient $D_i$ of Eq.\ (\ref{difchanel}) depends on the capillary contrast of the system. For very high capillary contrasts ($\eta_{A}>0.8$) the coupling coefficient can be taken as $D_i\simeq 0$, obtaining then the decoupled state observed numerically. On the other hand, for lower capillary contrasts ($\eta_{A}\simeq 0.72$) the variations of $D_{i}$ occur at scales larger than the correlation length and $D_{i}\simeq D$ can be taken as constant, obtaining then the regime described by $z\simeq 2$.
\begin{table*}[!]\caption{Complete set of scaling exponents for forced-flow imbibition when the capillary contrast $\eta_{A}$ is increased.}\label{tab} 
\centering
 \begin{ruledtabular} 
 \begin{tabular}{c||c c c c c c c c|}
   $\eta_{A}$ & $\alpha$ & $\alpha_{loc}$ & $\alpha_{s}$ & $z$ & $\beta$ & $\beta^{*}$ & Scaling class& \\
   \hline 
   $0.66$ & $1.5$ & $1$ & $1.2$ & $3$ & $0.49$ & $0.12$ & Superroughness & Non-local description \\
   \hline
   $0.72$ & $1.1$  & $0.6$  & $0.6$  & $2.2$ & $0.52$ & $0.26$ & Intrinsic A.& \\
   $0.74$ & $1.85$ & $0.55$ & $0.55$ & $3.7$ & $0.51$ & $0.35$ & Intrinsic A.&  Local \\
   $0.78$ & $2.5$  & $0.5$  & $0.5$  & $5.1$ & $0.5$  & $0.41$ & Intrinsic A.&  Description \\
   $0.82$ & $\infty$ & $0.5$ & $0.5$ & $\infty$ & $0.5$ & $0.5$ & Decoupled state & \\   
  \end{tabular}
  \end{ruledtabular}   
\end{table*}

\subsection{Spontaneous Imbibition}
In Fig.\ \ref{fig:IntA}b there is plotted four interface profiles evaluated at the same time but different capillary contrast. The velocity profiles of the interface are  shown in Fig.\ \ref{fig:Vmit}b. Since now there is not any imposed velocity, the velocity of the interface follows the Washburn's law in each channel of different noise value.\\
When we increase the parameter $\eta_{A}$ a transition to a decoupled state ($\beta=\beta^{*}=0.5$, $z=\infty$) also appears in the case of spontaneous imbibition. However, there is an important difference with the forced-flow case. As before, we calculate the local width in a small window size $\ell=L/128$ for different capillary contrasts (Fig.\ \ref{fig:wA}b). We observe that now, the local growth exponent $\beta^{*}$ changes suddenly to $\beta^{*}=0.5$, indicating that the interface advances completely decoupled. The transition to the decoupled state can be now discontinuous in time for each value of the capillary contrast. It seems that there exists a length $\ell_{d}(\eta_{A})$ above which the fluctuations become decoupled. Therefore, for a given value of the capillary contrast, the clusters of interface with a size $\ell_{d}$ will become decoupled between each other at the time $t_{d}\sim\ell_{d}^{z}$, when correlation length $\ell_{c}\sim t^{1/z}$ reaches the length $\ell_{d}$. Above $t_{d}$, the local description is not valid anymore.

\section{Conclusions} \label{Sec:Conclusions}
By means of numerical integrations of a phase-field model we find that there are strong differences between the dynamics of fluctuations in the cases of low and high capillary contrast with columnar disorder. Although these differences also exist in a case of quenched noise composed of squares \cite{intrinsic-superrough}, they are more dramatic when the quenched noise is of columnar type. It turns out that the persistence of the noise forces the interface to adopt the same geometry, changing the nature of the interface motion.\\
In a low capillary contrast case, interfaces are superrough with a dynamics dominated by surface tension with exponents $z=3$ when the velocity is nearly constant or $z=4$ when the velocity varies with Washburn's law. Furthermore, as interfaces are smooth and can be linearized around their mean value, a simple non local model for the interfacial evolution can be used to explain the observed dynamical scaling of fluctuations. On the other hand, when the capillary contrast is increased,  interfaces are sharper and the correlation with disorder is more evident. The observed dynamical scaling corresponds then to an anomalous scaling description with a clear temporal shift at the power spectrum, and a subdiffusive behavior with dynamical exponents ranging from $z=2$ to $z\to \infty$, depending on the strength of capillary forces. A  prominent point to remark is that this behavior can be explained by a local model made of coupled channels with a fluctuating force  following Darcy's law. One can interpret that in the high contrast case, the columnar disorder  induces the existence of channels with more or less coupling, eliminating completely the non local character of imbibition in homogeneous geometries. Finally, the difference between forced-flow and spontaneous imbibition has also been elucidated in the high capillary contrast regime. While in the forced-flow case the interface gets completely decoupled above a critical capillary contrast, in spontaneous imbibition the same decoupled state can be suddenly achieved for a given value of capillary contrast.

\section{Acknowledgments}
We acknowledge financial support from DGI of the Ministerio de Educaci\'on y Ciencia (Spain), projects FIS2006-12253-C06-04 and FIS2006-12253-C06-05.


\begin{thebibliography}{100}
%
\bibitem{Sahimi-95} M. Sahimi, {\it Flow and transport in porous media
and fractured rock}, John Wiley and Sons, New York (1995).

\bibitem{AL04}{M. Alava, M. Dub\'e, and M. Rost, \emph{Adv. Phys.} \pmb{53}, 83 (2004).}

\bibitem{Barabasi-Stanley} A.-L. Barab\'asi and H.E. Stanley, {\it
Fractal Concepts in Surface Growth}, Cambridge University Press, Cambridge
(1995).

\bibitem{Family-Vicsek-1985}
F. Family and T. Vicsek, J. Phys. A {\bf 18}, L75 (1985).

\bibitem{RA00}{J.J. Ramasco, J.M. L\'opez, and M.A. Rodr\'iguez, \emph{Phys. Rev. Lett} \pmb{84}, 2199 (2000).}

\bibitem{LE93}{H. Leschhorn and L.-H. Tang, \emph{Phys. Rev. Lett.} \pmb{70}, 2973 (1993).}

\bibitem{LO98}{J.M. L\'opez and J. Schmittbuhl, \emph{Phys. Rev. E} \pmb{57}, 6405 (1998).}

\bibitem{MA97}{J. Maunuksela, M. Myllys, O.-P. K\"ahk\"onen, J. Timonen, N. Provatas, M.J. Alava, and T. Ala-Nissila, \emph{Phys. Rev. Lett.} \pmb{79}, 1515 (1997).}

\bibitem{SO02b}{J. Soriano, J.J. Ramasco, M.A. Rodr\'iguez, A. Hern\' andez-Machado, and J. Ort\'in, \emph{Phys. Rev. Lett.} \pmb{89}, 026102 (2002).}

\bibitem{SO03}{J. Soriano, J. Ort\'in, and A. Hern\' andez-Machado \emph{Phys. Rev. E} \pmb{67}, 056308 (2003).}

\bibitem{SO05}{J. Soriano, A. Mercier, R. Planet, A. Hern\' andez-Machado, M.A. Rodr\'iguez, and J. Ort\'in, \emph{Phys. Rev. Lett.} \pmb{95}, 104501 (2005).}

\bibitem{Buldyrev-92-I} S. V. Buldyrev, A.-L. Barab\'asi, F. Caserta, S. Havlin,
H. E. Stanley, and T. Vicsek,
Phys. Rev. A, {\bf 45}, R8313 (1992).

\bibitem{Horvath-95} V.K. Horv\'ath and H. E. Stanley, Phys. Rev. E, {\bf 52},
5166 (1995).

\bibitem{Kwon-96} T. H. Kwon, A. E. Hopkins, and S. E. O'Donnell, Phys. Rev. E,
{\bf 54}, 685 (1996).

\bibitem{Zik-97} O. Zik, E. Moses, Z. Olami, and I. Webman, Euro. Phys. Lett.,
{\bf 38}, 509 (1997).

\bibitem{Balankin-2000} A. S. Balankin, A. Bravo-Ortega, and D. Morales, Phylos.
Mag. Lett., {\bf 80}, 503 (2000).

\bibitem{Rubio-89} M.A. Rubio, C.A. Edwards, A. Dougherty, and J.P. Gollub,
Phys. Rev. Lett.,
{\bf 63}, 1685 (1989).

\bibitem{Horvath-91} V.K. Horv\'ath, F. Family, and T. Vicsek, J. Phys. A, {\bf
24}, L25
(1991).

\bibitem{He-92} S. He, G.L.M.K.S. Kahanda, and P.-Z. Wong, Phys.
Rev. Lett.,
{\bf 69}, 3731 (1992).

\bibitem{SO02}{J. Soriano, J. Ort\'in, and A. Hern\' andez-Machado \emph{Phys. Rev. E} \pmb{66}, 031603 (2002).}

\bibitem{Geromichalos-2002} D. Geromichalos, F. Mugele, and S. Herminghaus,
Phys. Rev. Lett., {\bf 89}, 104503 (2002).

\bibitem{DU99}{M. Dub\'e, M. Rost, K.R. Elder, M. Alava, S. Majaniemi, and T. Ala-Nissila, \emph{Phys. Rev. Lett.} \pmb{83}, 1628 (1999).}

\bibitem{EL01}{K.R. Elder, M. Grant, N. Provatas, and J.M. Kosterlitz, \emph{Phys. Rev. E} \pmb{64}, 021604 (2001).}

\bibitem{Aurora-EPL-01} A. Hern\'andez--Machado, J. Soriano, A.M.
Lacasta, M.A. Rodr\'{\i}guez, L. Ram\'{\i}rez--Piscina, and J. Ort\'{\i}n,
Europhys. Lett.
{\bf 55}, 194 (2001).

\bibitem{HM03}{A. Hern\'andez-Machado, A.M. Lacasta, E. Mayoral, and E. C. Poir\'e, \emph{Phys. Rev. E}, \pmb{68}, 046310 (2003).}

\bibitem{LA05}{T. Laurila, C. Tong, I. Huopaniemi, S. Majaniemi, and T. Ala-Nissila, \emph{Eur. Phys. J. B} \pmb{46}, 553 (2005).}

\bibitem{BA89} G. Barton {\it
Elements of Green's functions and propagation: potentials, diffusion and waves}, Oxford University Press, Oxford 
(1989).

\bibitem{PA03}{E. Paun\'e and J. Casademunt, \emph{Phys. Rev. Lett.} \pmb{90}, 144504 (2003).}

\bibitem{Pradas07} M. Pradas, J. M. L\'opez, and A. Hern\'andez--Machado,
Phys. Rev. E {\bf 76}, 010102(R) (2007).

\bibitem{intrinsic-superrough}
M. Pradas and A. Hernandez-Machado, Phys. Rev. {\bf E74} 041608
(2006).

\bibitem{LO95}{J.M. L\' opez and M.A. Rodr\'iguez, \emph{Phys. Rev. E} \pmb{52}, 6442 (1995).}

\end{thebibliography}
\end{document}